\documentclass[prl,twocolumn,amsmath,amssymb,amsfonts,superscriptaddress,longbibliography]{revtex4-1}

\usepackage{graphicx}
\usepackage{hyperref}
\hypersetup{colorlinks=true}

\begin{document}

\title{Generation of Radio Frequency Radiation by Femtosecond Filaments}


\author{Travis Garrett}
\author{Jennifer Elle}
\author{Michael White}
\author{Remington Reid}
\affiliation{Air Force Research Laboratory, Directed Energy Directorate, Albuquerque, NM 87123, USA}
\author{Alexander Englesbe}
\affiliation{Naval Research Laboratory, Plasma Physics Division, Washington, DC 20375, USA}
\author{Ryan Phillips}
\author{Peter Mardahl}
\author{Erin Thornton}
\author{James Wymer}
\affiliation{Air Force Research Laboratory, Directed Energy Directorate, Albuquerque, NM 87123, USA}
\author{Anna Janicek}
\author{Oliver Sale}
\affiliation{Leidos Innovations Center, Albuquerque, NM 87106, USA}
\author{Andreas Schmitt-Sody}
\affiliation{Air Force Research Laboratory, Directed Energy Directorate, Albuquerque, NM 87123, USA}

\begin{abstract}

Recent experiments have shown that femtosecond filamentation plasmas 
generate ultra-broadband radio frequency radiation (RF). 
We show that a combination of plasma dynamics is responsible for the RF: 
a plasma wake field develops behind 
the laser pulse, and this wake excites 
(and copropagates with) a surface wave on the plasma column. 
The surface wave proceeds to detach from the end of the plasma 
and propagates forward as the RF pulse. 
We have developed a four stage 
model of these plasma wake surface waves 
and find that it accurately predicts the RF 
from a wide range of experiments, including both 800 nm 
and 3.9 $\mu$m laser systems.

\end{abstract}

\maketitle

\newcommand{\be}{\begin{equation}}
\newcommand{\ee}{\end{equation}}
\newcommand{\dnu}{\scriptstyle\mathcal{W}}


The development of chirped pulse amplification \cite{strickland1985compression} 
has enabled an array of exotic physics \cite{faure2004laser,liu1997laser,di2012extremely}, 
including atmospheric filamentation \cite{braun1995self}. 
Filamentation in turn produces interesting 
and useful effects, 
from supercontinuum generation \cite{brodeur1999ultrafast,kandidov2003self},
broadband THz radiation \cite{amico2008forward,andreeva2016ultrabroad}, and RF pulses, 
which have recently been explored in great detail  
\cite{forestier2010radiofrequency,englesbe2018gas,
janicek2020length,mitrofanov2021coherently,englesbe2021ultrabroadband,
mitrofanov2021polarization}.
We have discovered that a novel combination 
of a plasma wake field and a surface wave 
is responsible for this filament RF. 
Femtosecond lasers are becoming 
important spectroscopic tools 
(both in the lab \cite{kampfrath2013resonant} 
and for remote sensing \cite{labutin2016femtosecond}), 
and the added production of pulsed RF 
promises to expand the utility of these systems 
\cite{park2016perspective,doll2013adiabatic,capmany2007microwave},
and to provide insight  
into the physics of filamentation. 

During typical $\lambda=$ 800 nm filamentation 
Kerr self-focusing and plasma defocusing are in rough 
balance, resulting a plasma column with a density of  
 $n_e \sim 10^{22}$ m${}^{-3}$ 
\cite{couairon2007femtosecond,chin2010femtosecond}. 
This electron density corresponds to a plasma 
frequency of $\omega_{pl} \simeq 6 \times 10^{12}$ rad/s,
which is closely linked to 
the generation of THz radiation.
In the single color THz theory of \cite{amico2008forward}
the freshly ionized electrons receive a ponderomotive push
from the laser pulse, 
thus exciting a coherent longitudinal 
current $I_z$ in the plasma (which is quickly 
damped by the high collision frequency).
This short current pulse translates at $c$ behind the 
laser pulse, thereby producing a conical shell 
of radially polarized THz radiation. 

The RF resembles the THz radiation in some ways, 
and differs critically in others.
It has a similar conical shell spatial profile and radial polarization, 
which indicates that a GHz scale longitudinal current pulse 
is also generated within the plasma.
However it also has a well defined broadband peak in the 5-15 GHz range 
(and is not the tail of the THz radiation 
\cite{englesbe2018gas,englesbe2021ultrabroadband,mitrofanov2021coherently}) 
and the RF amplitude grows strongly with decreasing pressure  
\cite{englesbe2018gas,mitrofanov2021coherently}, 
in contrast to the THz which falls off sharply \cite{rodriguez2010scaling}. 
A distinct physical mechanism thus must be responsible for the RF.

Exploratory Particle In Cell (PIC) simulations 
\cite{birdsall2004plasma,peterkin2002virtual,
taflove2005computational,yee1966numerical,
boris1970relativistic,villasenor1992rigorous,
berenger1994perfectly} 
(see section $1$ in \cite{supplementary_ref_1} for more details)  
provided the key insights that explain the RF generation.
They revealed that a hot shell of electrons expands off the plasma surface 
into the surrounding atmosphere over roughly 50 ps
(as anticipated by \cite{zhou2011measurement}), 
and that the corresponding current density $J_r$ depends strongly
on the electron-neutral collision frequency.
Subsequent simulations demonstrated that this Plasma 
Wake field (PW) excites a Surface Wave (SW) 
on the outer boundary of 
the plasma (namely, a long wavelength 
Surface Plasmon Polariton (SPP)\cite{pitarke2006theory,maier2006terahertz}, 
with a speed approaching $c$).
Finally the SW was found to effectively detach from the end of the plasma, 
with most of the energy being then converted into a broadband RF pulse.

The PWSW numerical model is split into 4 stages, 
as length scales spanning 7 orders of magnitude need to be resolved.
First the initial electron distribution function 
$\mathcal{N}_e$ is determined.
In a typical $\lambda = 800$ nm experiment 
a linearly polarized, single color, 40 mJ 
pulse is focused into a quartz cell with a ${f}/60$ mirror, 
which sets the beam waist $\omega_0$ to 30 $\mu$m
(10 mJ and ${f}/15$ focusing are common for $\lambda =$ 3.9 $\mu$m).
The resulting Keldysh parameter 
$\gamma = \omega_{laser} \sqrt{2 m_e I_{pot}}/(q_e E)$ 
is close to 1, so we use the general ionization rate $\dnu$ 
from \cite{popruzhenko2008strong}, 
which is supported by experimental data \cite{sharma2018measurements}
near atmospheric pressure
(other ionization mechanisms will likely be needed   
for 10 $\mu$m lasers \cite{tochitsky2019megafilament}).
A laser intensity of $I \sim 5\times10^{17}$ W/m${}^2$ 
generates enough plasma at 1 atm for intensity clamping \cite{couairon2007femtosecond}
to occur (with $\dnu$ $\sim$ $10^{12}$ s${}^{-1}$),  
which leads to an estimated plasma length $L_{pl}$ of $25$ cm, 
and a radius $r_{pl}$ of $\sim$ $1$ mm
(see section 2 of \cite{supplementary_ref_1} for more details). 
At 1 Torr $I$ increases to $\sim 2\times10^{18}$ W/m${}^2$, 
which ionizes much of the O${}_2$, and 
the plasma is reduced to $L_{pl} \simeq 13$ cm 
and $r_{pl} \simeq 0.5$ mm.
In turn, the lower energy 3.9 $\mu$m laser plasmas  
are shorter, with $L_{pl}$ ranging from about $1$ to $2$ cm.

These estimates match well (within 25\%) with photographic 
measurements of the plasma taken 
at different pressures (available in 
section 2 of \cite{supplementary_ref_1}).
Simulations of filamentation with these parameters have also been 
performed using a Unidirectional Pulse Propagation Equations (UPPE) 
solver \cite{kolesik2004nonlinear}.
These agree with the overall dimensions of the 
plasma column, but also indicate that there is rich fine scale 
structure throughout the plasma 
(as expected as we are in the multifilamentary regime
\cite{berge2004multiple,ettoumi2015laser,ettoumi2015spin}). 
For this letter we simplify and do not include 
fine scale plasma perturbations. 
The plasma column is approximated as having 
a constant density central core 
due to intensity clamping,
and then a 100 $\mu$m radial ramp down to 
zero density at the outer radius $r_{pl}$.

The velocity distribution is determined by our PIC code: 
given $E(\vec{x},t)$ the air is ionized 
at the $\dnu$ rate \cite{popruzhenko2008strong}
and the new electrons 
are accelerated through the remainder of the pulse 
\cite{boris1970relativistic}
(code to perform these calculations is included 
in \cite{supplementary_ref_1}).
In general the velocity distribution 
is influenced by both the details of strong field 
ionization near $\gamma = 1$ (e.g. \cite{corkum1993plasma}),
and by the deformation of the laser pulse 
during filamentation.
For this work we further simplify and assume that 
the electrons are ionized with zero initial velocity and 
are then accelerated by the remainder 
of a Gaussian pulse (with $\hat{x}$ polarization 
and propagating in $+z$). 
As a whole the initial $\mathcal{N}_e$ is highly non-Maxwellian, 
and at 100 Torr has a peak kinetic energy $K_{tail} \simeq$ 5 eV, 
and an average kinetic energy $K_{avg} \simeq 0.6$ eV, 
while these increase at 1 Torr to $K_{tail} \simeq 16$ eV and $K_{avg} \simeq 2$ eV.
For 3.9 $\mu$m lasers the kinetic energy is $\sim 25$ times larger, 
as the laser intensities are  
comparable and the energy scales as $\lambda^2$.


We next consider the evolution of the plasma column. 
We construct thin transverse slices of plasma 
given $\mathcal{N}_e$, 
with periodic boundary conditions used in the 
longitudinal direction $\hat{z}$
(which is valid to leading order as the electron velocities 
are a small fraction of $c$)
and simulate the radial evolution with our PIC code.
The Debye length is quite small: $\lambda_{Debye} \simeq 10$ nm, 
so we use an energy conserving method \cite{pointon2008second}
to calculate the Lorentz force.
The electron-neutral elastic collision frequency 
$\nu_{eN}$ depends on the cross sections for O${}_2$ and N${}_2$,
which for our energies is roughly 10 \AA${}^2$ \cite{itikawa2006cross}.
In turn the electron-ion momentum-transfer collision frequency is 
given by $\nu_{ei} = 7.7{\times}10^{-12} n_e \ln ( \Lambda_C ) / K_{eV}^{3/2} $,
where $\Lambda_C = 6 \pi n_e \lambda_{Debye}^3$ \cite{thorne2017modern}. 
The resulting radial current density
$J_r$ and electron density $n_e$ 
are then recorded as functions of radius and time 
(more detail can be found in section 3 of \cite{supplementary_ref_1}).
These are well resolved with a mesh resolution 
of $\Delta x = \Delta y =$ 2 $\mu$m, 
and a macro-particle weight of 
${\sim}10$ on the outer edge of the plasma.

\begin{figure}[h!]
  \centering
    \includegraphics[width=0.5\textwidth]{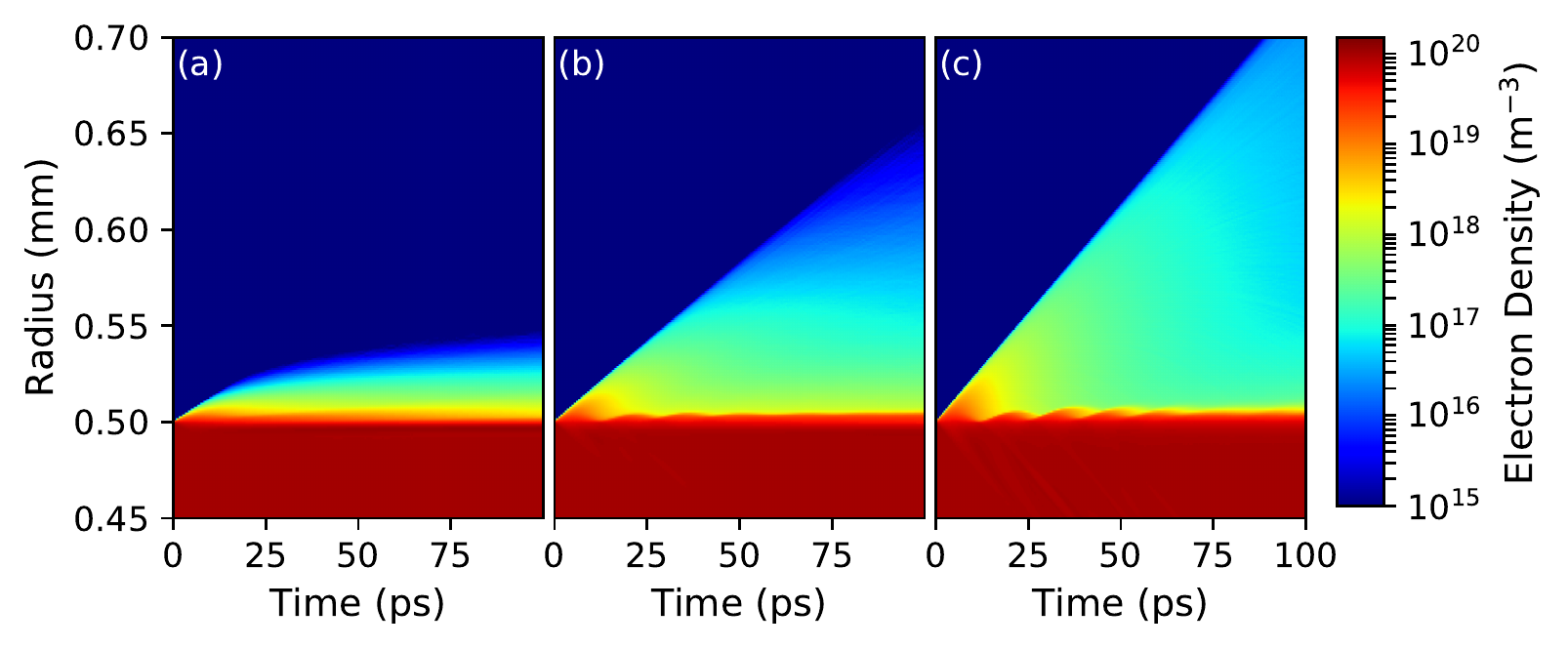}
      \caption{Log plots of the electron density $n_e(r,t)$ 
      showing the plasma wake variation at pressures 
      of 100 (a), 10 (b), and 1 Torr (c) 
      for a $\lambda = 800$ nm laser pulse.
      The plasma outer radius has been set to $r_{pl}=0.5$ mm 
      for all 3 pressures for comparison purposes.} 
      \label{log_ne_plots}
\end{figure}

\begin{figure}[h!]
  \centering
    \includegraphics[width=0.5\textwidth]{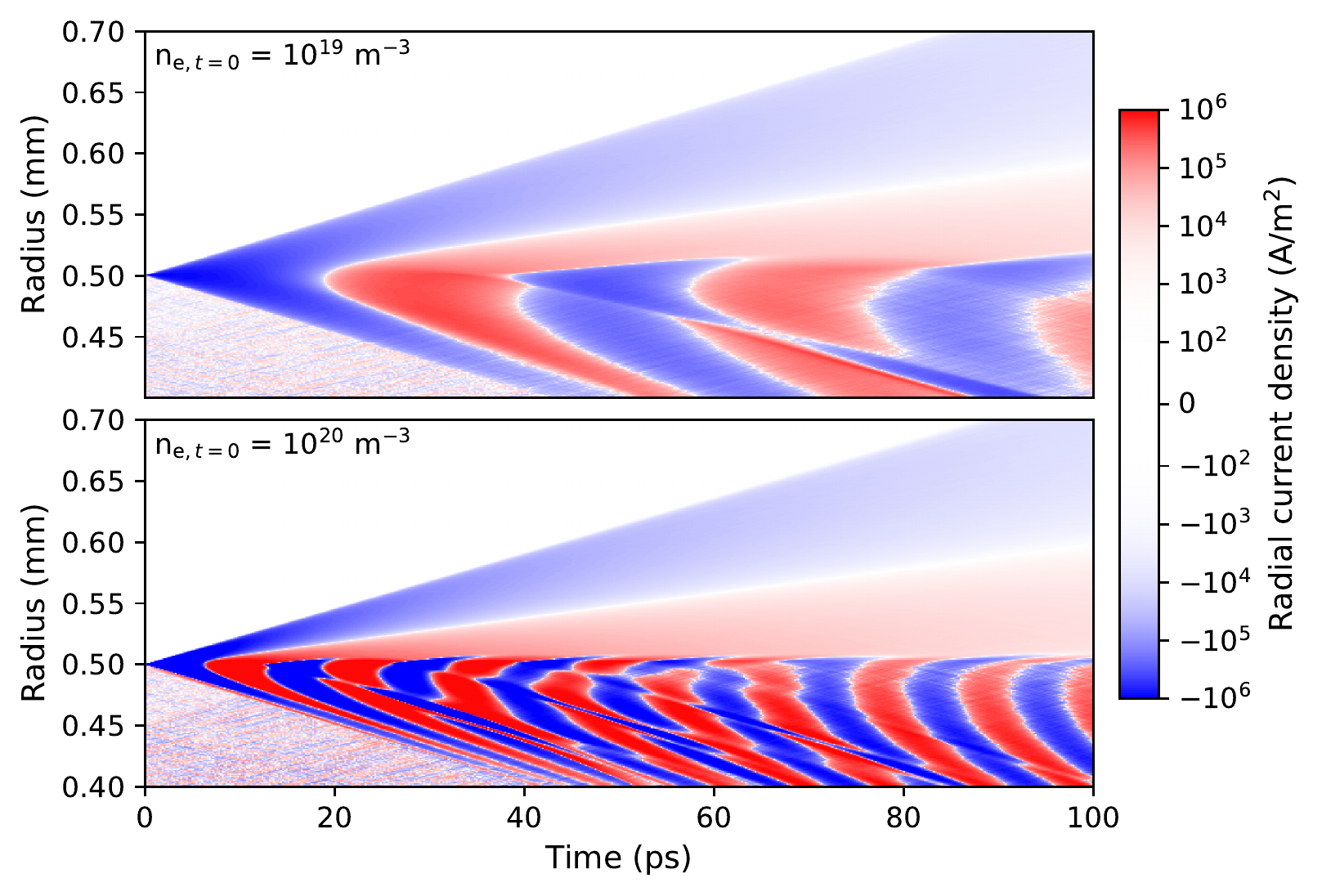}
      \caption{Log plots of the radial current density $J_r(r,t)$ 
at 1 Torr (800 nm source), 
as it evolves in a transverse slice of plasma. 
The initial density at the edge of the plasma has been given 
a step function profile, with a value of $n_e = 10^{19}$ m${}^{-3}$  
in the top plot and $n_e = 10^{20}$ m${}^{-3}$ on the bottom. 
The profiles of the plasma wakes outside the column ($r > r_{pl} = 0.5$ mm) 
are quite similar, while at the surface
SPPs are excited at roughly
${{\sim}}20$ and ${{\sim}}60$ GHz respectively, along with 
higher frequency volume plasmons in the interior.
} 
      \label{log_jr_plots}
\end{figure}

The electron number density for $\lambda=800$ nm simulations of the 
PW at 100, 10, and 1 Torr can be seen in Fig.~\ref{log_ne_plots}. 
The outer edge of the plasma at $t=0$ has a  
simplified step function profile, 
with $n_e = 10^{20}$ m${}^{-3}$
at radius $r_{pl} = 0.5$ mm.
Thus, in addition to the PW being launched off of the edge of this plasma, 
a coherent radial plasma frequency oscillation of ${\sim} 90$ GHz 
is stimulated in the interior \cite{dawson1959nonlinear} and a SPP 
at ${\sim} 63$ GHz is excited on the surface 
\cite{pitarke2006theory,maier2006terahertz,sakai2018wave}. 
The PW that expands out into the neutral atmosphere ($r > r_{pl}$) 
is insensitive to the density 
of the outer edge of the plasma (see Fig.~\ref{log_jr_plots}), 
as opposed to the excited surface and volume plasmons.  

Analytic approximations provide a useful complement to the simulations.
Consider the late time electron density at a small 
distance $r_{\Delta}$ off the surface of the plasma at radius $r_{pl}$.
Using Gauss's law and approximating $\ln( ( r_{pl} + r_{\Delta} ) / r_{pl} )$ 
as $r_{\Delta} / r_{pl}$ for $r_{\Delta} \ll r_{pl}$ 
we find the that the number of electrons per unit 
length $n_L$ that escape to $r_{\Delta}$ scales as:
\be
n_L(r_{\Delta}) \simeq \frac{2 \pi \epsilon_0 K_{eV} }{q_e} \frac{r_{pl}}{r_{\Delta}}, 
\label{nL_eq}
\ee
where $K_{eV} \simeq K_{tail}$ is the kinetic energy 
of the hot electrons that lead the escape.
The radial electrostatic field $E_{r,stat}$ that corresponds to $n_L$ has a 
simple form: $E_{r,stat}(r_{\Delta}) = K_{eV} / r_{\Delta}$.
At $\lambda=800$ nm and 1 Torr 
$K_{tail} \simeq 16$ eV and $r_{\Delta} \simeq$ 60 $\mu$m, 
giving $E_{r,stat} \simeq 2.7 \times 10^5$ V/m.
This allows for an estimate of the PW 
evolution timescale: 
an electron with energy $K_{tail} = 16$ eV 
in an electric field of this magnitude  
follows a parabolic trajectory over 100 ps. 
This compares well with the spread 
of electrons seen in Fig.~\ref{log_ne_plots}. 
The magnitude of the radial current density can likewise be approximated as:
\be
J_r \simeq v_{eff}(t) \frac{q_e n_L}{2 \pi r_{pl} r_{\Delta} }; v_{eff}(t) = \min \begin{cases} v_{tail} \\ v_{diff}(t) \end{cases}
\label{Jr_eq}
\ee
where the effective velocity $v_{eff}(t)$ is the minimum of the tail 
velocity and the Fickian diffusion speed 
$v_{diff}(t) = (K_{tail} /(m_e \nu t))^{1/2}$.
At 1 Torr this gives $J_r \simeq 1\times10^5$ A/m${}^{2}$.
Finally we note that caution is needed for 
small $r_{\Delta}$ values: Eq.s \eqref{nL_eq} and \eqref{Jr_eq}
only hold up to densities that are comparable to the original plasma 
edge density.

 
We next switch from these transverse PIC simulations  
to a continuum FDTD-Drude model in a 2D axisymmetric coordinate system 
\cite{taflove2005computational} that spans the length of the plasma.
The Drude model: 
$\partial_t \vec{J}_{pl} + \nu_{pl} \vec{J}_{pl} = \epsilon_0 \omega_{pl}^2 \vec{E}$ 
is integrated into the code 
via an auxiliary differential equation \cite{okoniewski1997simple}. 
A mixed resolution of $\Delta r =$ 2 $\mu$m and $\Delta z =$ 50 $\mu$m
suffices to resolve the SWs. 
The PW current profile $J_r$ is driven across 
the surface of the plasma column 
in the longitudinal direction at the speed of light. 
In this work only the external radial currents $J_{r > r_{pl}}$ 
are used to drive the SW and subsequent RF
(see section 4 of \cite{supplementary_ref_1} for more detail).
The higher frequency volume plasmons can also 
excite waves in neighboring lower density plasma, but we 
suspect that including fine scale plasma structure 
will act as surface roughness \cite{kretschmann1972angular}
and cause these to continuously detach. 

As expected the PW current  
excites a broadband surface wave with similar frequency content.
The frequency $\omega$ of the wave is considerably lower than
the limit SPP frequency 
and it thus travels at velocity 
$c ((2 \omega^2 - \omega_{pl}^2)/(\omega^2 - \omega_{pl}^2))^{1/2}$
which approaches the speed of light.
The resulting phase-matched copropagation leads to  
a steady growth in the surface wave intensity 
(see also \cite{smith1953visible,dahan2020resonant}).

The SW is well approximated by the Sommerfeld-Goubau  
\cite{sommerfeld1899ueber,goubau1950surface} 
solution for SPPs on a cylinder of finite conductivity. 
The external radial component of the electric field $E_r$ has the form: 
\be
E_{r}( r, z, t ) = -\frac{\pi r_{pl} E_0 }{2 r_{outer}} e^{i(\omega t - h z)} H^{(1)}_1 (r \sqrt{k^2 - h^2} ) ,
\label{SG_SW}
\ee
for waves of frequency $\omega$ propagating in the $+z$ direction,
with an amplitude of $E_0$ at the surface.
The complex $h$ wavenumber describes the SW wavelength and 
attenuation length scale, 
$k$ is the free space wavenumber $k=\omega / c$, 
$H^{(1)}_1$ is a Hankel  
function of the first kind with order 1, 
and $r_{outer} = 1/|\sqrt{k^2 - h^2}|$.
For radii $r_{pl} < r < r_{outer}$ 
the Hankel function with complex argument 
is approximately $H^{(1)}_1 \simeq -2 r_{outer}/(\pi r)$, 
and for larger radii it falls off exponentially 
(as is typical for SPPs).

Recent work \cite{jaisson2014simple,mendoncca2019electromagnetic} 
has simplified the calculation of $h$ 
through the use of the Lambert $W$ function.
Given a typical plasma conductivity $\sigma$ a 
complex $\xi$ can be defined:
\be
\xi = -\frac{k e^{-2 \gamma_{euler}}}{2} \sqrt{ \frac{ \epsilon_0 \omega }{ 2 \sigma } } r_{pl} (1+i),
\ee
(where $\gamma_{euler}$ is the Euler-Mascheroni constant), 
and $h$ is then given by:
\be
h = \sqrt{ k^2 + \frac{4 \xi e^{-2 \gamma_{euler}}}{r_{pl}^2 W_1(\xi)} },
\ee
where the first branch of $W$ is used.
At 1 Torr the plasma has a radius $r_{pl} \simeq 0.5$ mm 
and conductivity 
$\sigma \simeq 2 \times 10^3$ S/m, and thus for 
a 10 GHz SW we find $h \simeq 210 + 3.2i$.
The dispersive SW thus travels at about $0.988 c$, 
attenuates by a factor of $1/e$ every $30$ cm 
(if not being actively excited), 
and has a $1/r$ transverse envelope out to 
$r_{outer} \simeq 2.4$ cm, 
all of which compares well with simulations.

\begin{figure}
  \centering
  \includegraphics[width=0.45\textwidth]{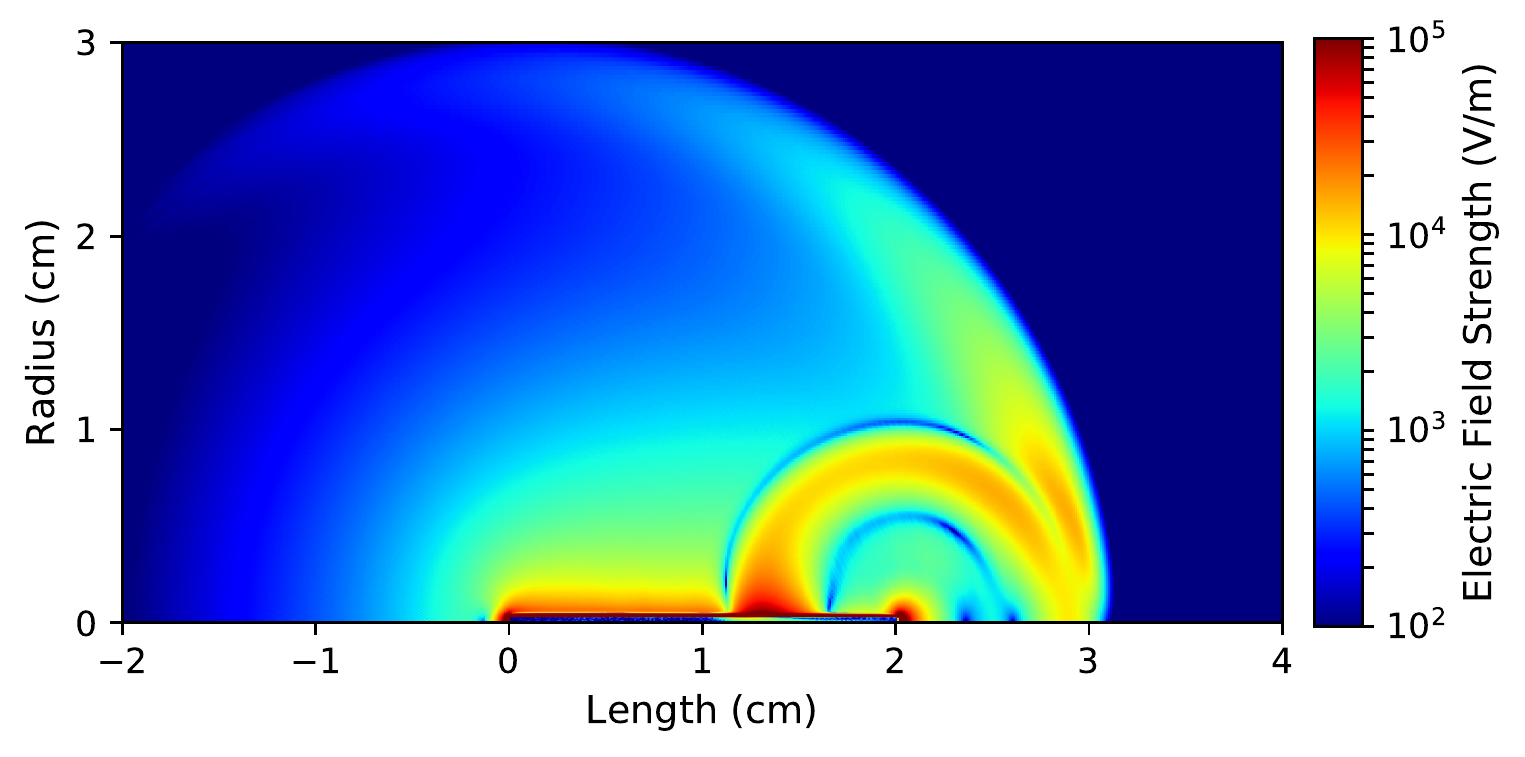} \\
  \caption{Log plot of electric field strength shortly after the detachment of a $\lambda=3.9$ $\mu$m, 10 Torr surface wave from a 2 cm plasma column, as 
simulated by the axisymmetric FDTD-Drude model.}
  \label{SW_39}
\end{figure}

We next consider the excitation of the surface wave
by the plasma wake currents.
An isolated Hertzian dipole with uniform current $I$ 
along an antenna of length $L$ will 
radiate electromagnetic waves at a wavelength $\lambda_{Dipole}$ 
with power:
$P_{Dipole} = (\pi\eta_0/3) I^2 L^2 / \lambda^2_{Dipole}$.
In our case we have a radial current distribution 
of approximate length $r_{\Delta}$
that is normal to a conducting cylinder of radius $r_{pl}$.
When there is a separation of length scales 
$r_{\Delta} \ll r_{pl} \ll \lambda_{SW}$ 
and an overall radial current of $I_r$ then the 
power that is primarily radiated into SWs is:
\be
P_{SW} \simeq \frac{\pi \eta_0}{60} I_r^2 \frac{r_{\Delta}^2}{r_{pl}^{5/3} \lambda^{1/3}_{SW}},
\label{power_SW}
\ee
(where we have simplified with $\lambda_{SW} = 2\pi/\operatorname{\mathbb{R}e}(h)$).
We note that the dimensionless term 
$r_{\Delta}^2/(r_{pl}^{5/3} \lambda^{1/3}_{SW})$ 
(compare to $L^2 / \lambda^2_{Dipole}$ for the Hertzian dipole),
has different asymptotic scalings in other length scale regimes.
Equation \eqref{power_SW} applies both to a stationary 
radial antenna and the $\pm z$ wavetrains it emits, and to our case, 
where the antenna translates at $c$ with its comoving SW pulse.

The total effective current $I_r(z)$ 
is found by integrating 
the current density $J_r$ from 0 to $z$, 
and absent dissipation the amplitude $E_0$ 
of the SW in \eqref{SG_SW} would grow linearly (for constant $J_r$). 
However there is dissipation due to the finite 
conductivity of the plasma, 
and the attenuation coefficient $\alpha = \operatorname{\mathbb{I}m}(h)$ 
is similar in magnitude to $1/L_{pl}$.
The SW thus grows linearly initially 
and then saturates after several $1/\alpha$, and the effective 
total radial current after propagating a distance $z$ is:
\be
I_r(z) = 2 \pi r_{pl} J_r \frac{1 - e^{-\alpha z}}{\alpha}.
\label{Ir_integrated}
\ee
We express the amplitude of the final SW in terms of 
its longitudinal current, as $I_z = 2 \pi r_{pl} E_0 / \eta_0$.
Integrating \eqref{SG_SW} to find the SW peak power 
and setting this equal to \eqref{power_SW} we find: 
\be
I_z(z) = \frac{2 \pi^2 r_{\Delta} J_r (1-e^{-\alpha z})}{\alpha(30 \ln (r_{outer}/r_{pl}))^{1/2}}
\left( \frac{r_{pl}}{\lambda_{SW}} \right)^{1/6}.
\label{final_Iz}
\ee
For the $\lambda=800$ nm laser this gives 
$I_z \simeq 0.6$ A at the end of the 
plasma at 1 Torr, a current of $I_z \simeq 0.2$ A at 10 Torr, 
and $I_z \simeq 0.05$ A at 100 Torr, which are close to the 
measured currents in the corresponding simulations.


When the surface wave reaches the end of plasma the majority 
of the energy detaches and propagates forward as the RF pulse, 
with the reflected wave decaying as it travels back down the plasma.
Fig.~\ref{SW_39} shows an example of this, 
where the SW has just
detached from a 2 cm plasma column produced by a
$\lambda=$ 3.9 $\mu$m laser source at 10 Torr.
In this way the system resembles a surface wave 
end-fire antenna \cite{andersen1967radiation,sporer201724}.  
The second half of the primarily one cycle RF pulse forms as the 
SW separates from the plasma, with some variation in the profile 
dependent on the abruptness of the detachment.
The recorded $I_z(z,t)$ currents from the SW simulation 
are used in a ${\sim}1$ m${}^2$ 
laboratory scale axisymmetric FDTD model, 
with resolution $\Delta r = \Delta z = 0.5$ mm. 
Plots of the resulting Fourier transformed RF field strengths 
at 10 Torr for 
both $\lambda = 800$ nm and $\lambda = 3.9$ $\mu$m
simulations can be seen in Fig.~\ref{Far_RF_1}
(see section 5 of \cite{supplementary_ref_1} for more detail).

\begin{figure}[h!]
  \centering
    \includegraphics[width=0.5\textwidth]{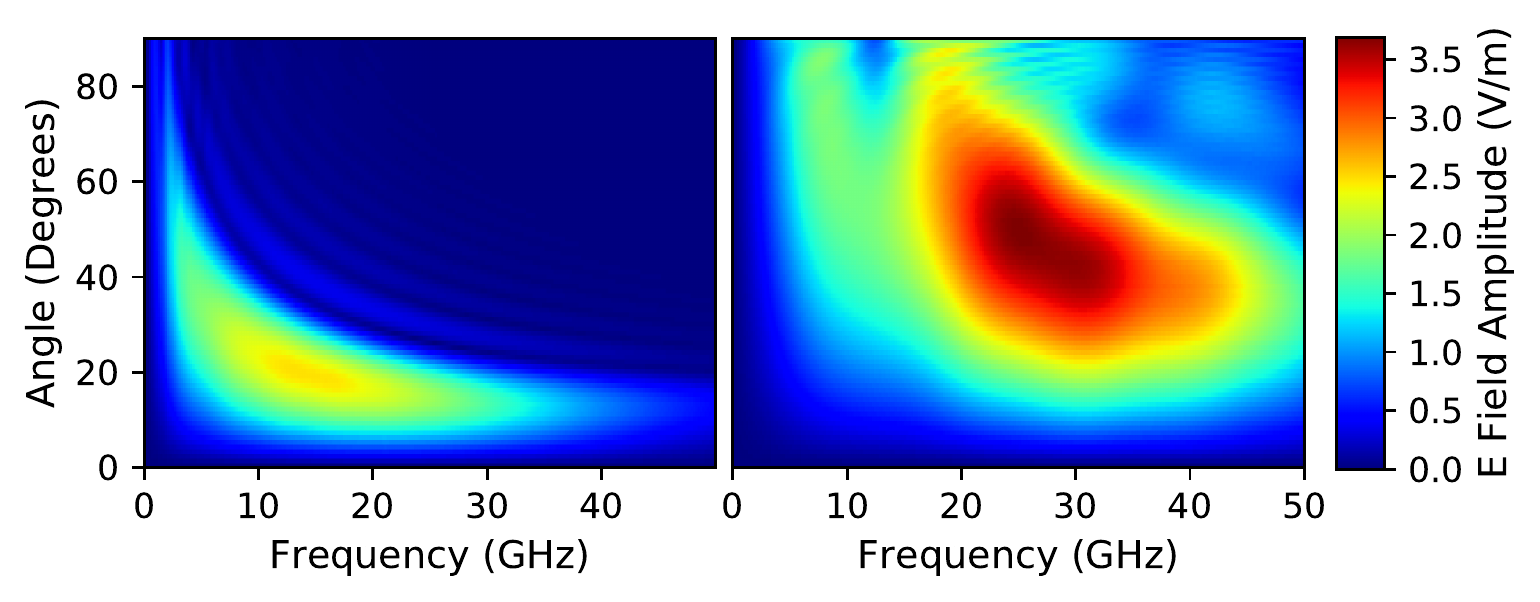}
      \caption{Plots of simulated electric field strength as a function of frequency 
        and angle for RF pulses from a 40 mJ, f${}_N=60$, 800 nm laser source 
        (left, \cite{englesbe2018gas}) and a 12 mJ, f${}_N=15$, 3.9 $\mu$m source 
        (right, \cite{mitrofanov2021coherently}), with both cases run at 10 Torr.  
        The numerical profiles and amplitudes 
        closely match the experimental results. 
  } 
      \label{Far_RF_1}
\end{figure}

\begin{figure}[h!]
  \centering
    \includegraphics[width=0.5\textwidth]{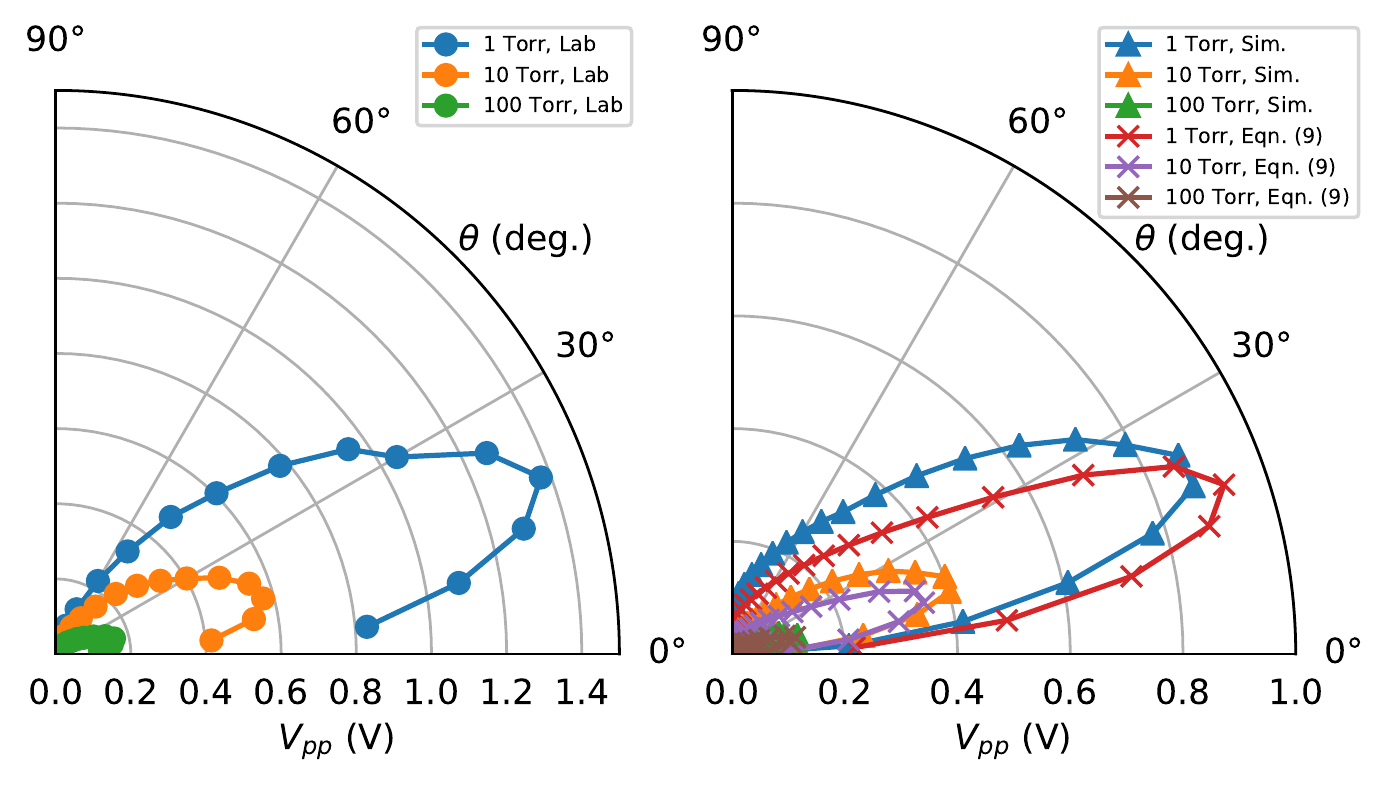}
      \caption{Polar plots of the detector peak to peak Voltage $V_{pp}$ 
              for the $\lambda=800$ nm laser source 
              as a function of angle and pressure as 
              measured in the lab (left), and as calculated 
              by both the simulation framework (right, triangles), 
              and by the associated analytic approximations 
              (right, $\times$ ticks) that culminate in \eqref{THz_eqn}.} 
      \label{Vpp_compare}
\end{figure}

To directly compare with lab results the simulated 
electric fields are converted to peak 
to peak voltages $V_{pp}$ via a frequency
dependent calibration factor  
$C_{E/V}(f_{GHz}) \simeq 50 + 10 f_{GHz}$
(which captures the horn antenna response \cite{englesbe2018gas}).
We adapt an equation for energy spectral density 
radiated per unit solid angle derived 
for THz radiation \cite{amico2008forward}: 
\be
\frac{d^2 U}{d\omega d\Omega} = \frac{|I_z(\omega)|^2 \sin^2 \theta }{4 \pi \epsilon_0 c (1-\cos \theta)^2}
\sin^2 \left( \frac{L_{pl} \omega}{2c} (1-\cos \theta) \right),
\label{THz_eqn}
\ee
to construct voltage profiles for our analytic approximations. 
The average amplitude of $I_z$ \eqref{final_Iz} 
over the length of the plasma is applied to a ${\sim}10$ GHz pulse, 
which is Fourier transformed for use with \eqref{THz_eqn}
(along with the calibration factor), and then 
converted back to an effective $V_{pp}$
(code that implements all of these 
analytic approximations is available in \cite{supplementary_ref_1}).

In general simulations with varying pressures,
focal geometries, and  
laser pulse energies and wavelengths 
all produce RF that nicely matches recent 
experimental results  
\cite{englesbe2018gas,janicek2020length,
mitrofanov2021coherently,englesbe2021ultrabroadband}.
The magnitude of the simulated RF is a bit lower than the 
lab data for the runs shown in Fig.~\ref{Vpp_compare},
but is within the errors for both the 
simulations and in the experimental data.
Subtle changes in Ti:Sapphire laser alignment 
generates $\pm 50\%$ amplitude variations at a given pressure 
(with the spatial and frequency content being much more stable).
It is suspected that the slight laser pulse variations 
are amplified during filamentation, 
leading to fine scale changes in the plasma radius, 
density, and electron velocities.
Plausible variations in the same parameters lead to 
comparable changes in the simulated RF amplitude as well.


We have demonstrated that a plasma wake field that 
excites a surface wave is responsible 
for the generation of filament RF.
The rich internal dynamics of the plasma 
merits further research: we expect that 
stochastic higher frequency RF 
may inform on the fine scale structure of the plasma and the 
nonlinear optics that generate it.
The ionization model needs to be expanded to include more physical effects
(e.g. \cite{tochitsky2019megafilament,corkum1993plasma,amini2019symphony}),  
and strong field ionization can be further explored with noble gas 
filament RF.
The plasma can also be driven into the  
strongly coupled regime with harder focusing, 
and we intend to model resulting RF 
with the use of \cite{baalrud2013effective,donko2009molecular}.
Finally we note that it may be possible 
to recreate the core mechanism 
-- an antenna that copropagates with the wave that it excites -- 
in other settings and thereby 
drive strong coherent radiation.

\begin{acknowledgments}
\vspace{4mm}

The authors thank Edward Ruden, Serge Kalmykov and Charles Armstrong 
for useful discussions,
and the Air Force Office of Scientific Research (AFOSR) for support
via Laboratory Tasks No. FA9550-18RDCOR0009 and No. FA9550-19RDCOR027.
This work was supported in part by high-performance computer time 
and resources from the DoD High Performance Computing Modernization Program.
Approved for public release; distribution is unlimited. Public Affairs
release approval No. AFRL-2020-0489.

\end{acknowledgments}

\bibliography{../bib_rf}

\end{document}